\documentclass[showkeys,aps,prd,eqsecnum,twocolumn,twoside]{revtex4-1} 

\usepackage{amsfonts,amsmath,amssymb,bm}

\usepackage[active]{srcltx}

\begin{document}

\title{Evolution of a dense neutrino gas in matter and electromagnetic field}

\author{Maxim Dvornikov$^{a,b}$}
\email{maxdvo@izmiran.ru}
\affiliation{$^a$N.~V.~Pushkov Institute of Terrestrial Magnetism,
Ionosphere and Radiowave Propagation (IZMIRAN), \\ 142190 Troitsk,
Moscow Region, Russia; \\
$^b$Institute of Physics, University of S\~{a}o Paulo, \\
CP 66318, CEP 05315-970 S\~{a}o Paulo, SP, Brazil}

\date{\today}

\begin{abstract}
We describe the system of massive Weyl fields propagating in a
background matter and interacting with an external electromagnetic
field. The interaction with an electromagnetic field is due to the
presence of anomalous magnetic moments. To canonically quantize
this system first we develop the classical field theory treatment
of Weyl spinors in frames of the Hamilton formalism which accounts
for the external fields. Then, on the basis of the exact solution
of the wave equation for a massive Weyl field in a background
matter we obtain the effective Hamiltonian for the description of
spin-flavor oscillations of Majorana neutrinos in matter and a
magnetic field. Finally, we incorporate in our analysis the
neutrino self-interaction which is essential when the neutrino
density is sufficiently high. We also discuss the applicability of
our results for the studies of collective effects in spin-flavor
oscillations of supernova neutrinos in a dense matter and a strong
magnetic field.
\end{abstract}


\date{\today}

\maketitle

\section{Introduction}

It is known that the neutrino interaction with external fields can
significantly influence the evolution of supernova
neutrinos~\cite{Raf96}. For example, the combined action of a
background matter and a magnetic field can cause the resonant
transition like, $\nu_\alpha^{-{}} \leftrightarrow
\nu_\beta^{+{}}$ (see, e.g., Ref.~\cite{LimMar88}), where the
indexes $\pm{}$ denote different helicity states. Hence the active
neutrinos of the flavor $\alpha$ can be converted into sterile
neutrinos of another flavor $\beta$.

Besides external fields, other factors, like the neutrino
self-interaction, can strongly influence neutrino oscillations. It
happens, e.g., in a supernova explosion, when the typical neutrino
luminosity can be $\sim
10^{51}\thinspace\text{erg/s}$~\cite{GiuKim07}. Accounting for the
average supernova neutrino energy $E_\nu \sim
10\thinspace\text{MeV}$, we get that at the distance about several
tens of kilometers from the protoneutron star surface the neutrino
number density still can be high enough for interactions between
neutrinos to be as important as the neutrino interaction with
external fields. This neutrino self-interaction leads to the
collective effects in neutrino oscillations.

For the first time the neutrino self-interaction was considered in
Ref.~\cite{Ful87}. Since then a lot of works on this subject has
been published (see, e.g., the recent review~\cite{DuaFulQia10}
and references therein). Note that in the majority of the studies
of collective effects in neutrino oscillations only the
combination of the interaction with a background matter and the
neutrino self-interaction was considered (see, e.g.,
Ref.~\cite{Sam93}). In the present work we shall generalize the
previous approaches for the description of collective neutrino
oscillations to include the interaction with an external
electromagnetic field since, as we mentioned above, the influence
of a strong magnetic field on the neutrino system evolution can be
also important.

Neutrinos can interact with an external electromagnetic field due
to the presence of anomalous magnetic moments. Note that the
structure of the magnetic moments is completely different for
Dirac and Majorana neutrinos (see, e.g., Ref.~\cite{FukYan03}).
Despite the fact that nowadays there is no universally recognized
confirmation of the nature of neutrinos~\cite{EllEng04}, in the
present work we shall suppose that neutrinos are Majorana
particles. Note that in various scenarios for the generation of
elementary particles masses, it is predicted that neutrinos should
acquire Majorana masses~\cite{MohSmi06}.

In the present work we shall resolve several important problems
for the physics of Majorana neutrinos interacting with external
fields. First, in Sec.~\ref{CLASS}, using the results of our
recent paper~\cite{Dvo11} we propose the classical field theory
treatment of massive Weyl fields propagating in background matter
and interacting with an external electromagnetic field. Then, in
Sec.~\ref{QUANT}, on the basis of the exact solution of the wave
equation for Weyl fields in a background matter we canonically
quantize these fields. In Sec.~\ref{MATTEMF}, in frames of our
method we re-derive the effective Hamiltonian for the description
of spin-flavor oscillations of Majorana neutrinos in matter and a
magnetic field. Finally, in Sec.~\ref{SI}, we apply the developed
formalism to get the contribution of the neutrino self-interaction
to the effective Hamiltonian. In Sec.~\ref{CONCL}, we summarize
our results.

\section{Classical field theory\label{CLASS}}

In this section we develop the classical field theory description
of the massive neutrinos eigenstates, which are supposed to be
Majorana particles, in matter and electromagnetic field. For this
purpose we derive the Hamiltonian for the system of two-component
Weyl spinors and show that the classical canonical equations are
equivalent to the wave equation for a Majorana neutrino in
external fields.

The wave equation for the neutrino mass eigenstates $\psi_a$,
propagating in background matter and interacting with an external
electromagnetic field, is known to have the form,
\begin{equation}\label{wepsi}
  (\mathrm{i} \gamma^\mu \partial_\mu - m_a) \psi_a -
  \frac{\mu_{ab}}{2} \sigma_{\mu\nu} F^{\mu\nu} \psi_b +
  g_{ab}^\mu \gamma_\mu \gamma^5 \psi_b = 0,
\end{equation}
where $m_a$ are the masses of the particles, $\gamma^\mu$,
$\gamma^5$, and $\sigma_{\mu\nu} = (\mathrm{i}/2)(\gamma_\mu
\gamma_\nu - \gamma_\nu \gamma_\mu)$ are the Dirac matrices. Note
that we will formulate the dynamics of the system~\eqref{wepsi} in
the mass eigenstates basis rather than in the flavor basis, as it
is usually done when neutrino oscillations are considered, since
only in the mass eigenstates basis one can distinguish between
Dirac and Majorana masses~\cite{Kob82}.

The interaction with matter is characterized by the external
fields $(g_{ab}^\mu)$, which, in principle, are nondiagonal in the
neutrino mass eigenstates basis. In general case the matrix
$(g_{ab}^\mu)$ is hermitian. However we shall discuss the
situation when the CP invariance is conserved. Despite a current
attempt to detect CP violating terms in the neutrino
sector~\cite{Abe11}, no definite results have been obtained yet.
In this case the vacuum mixing matrix is orthogonal and the matrix
$(g_{ab}^\mu)$ is symmetric. The zero component of this matrix,
$(g_{ab}^0)$, contains the effective potentials of the neutrino
interaction with non-moving and unpolarized matter, whereas the
vector components, $(\mathbf{g}_{ab})$, are the linear
combinations of the averaged matter velocity and the polarization.
The explicit form of these matrices and the details of the
statistical averaging can be found in Ref.~\cite{DvoStu02}.

Note that the vector term in the neutrino matter interaction $\sim
g_{ab}^\mu \gamma_\mu \psi_b$ is omitted in Eq.~\eqref{wepsi}
since it is washed out for Majorana neutrinos. The contribution of
the axial-vector interaction with matter to the wave
equation~\eqref{wepsi} $\sim g_{ab}^\mu \gamma_\mu \gamma^5
\psi_b$ is twice the analogous contribution for Dirac particles
since both neutrinos and antineutrinos equally interact with a
background matter (see, e.g., Ref.~\cite{GriStuTer05}).

Neutrinos can interact with an external electromagnetic field
$F_{\mu\nu} = (\mathbf{E},\mathbf{B})$ owing to the presence of
the anomalous magnetic moments $(\mu_{ab})$. It is know (see,
e.g., Ref.~\cite{PasSegSemVal00}) that the matrix $(\mu_{ab})$
should be hermitian and pure imaginary, i.e. $\mu_{ab} = -
\mu_{ba}$ and $\mu_{ab}^{*{}} = - \mu_{ab}$. We shall discuss the
situation when no admixture of sterile neutrinos is in the mass
eigenstates $\psi_a$. In this case the electric dipole moments are
equal to zero~\cite{PasSegSemVal00}.

We define the interaction with external fields in the mass
eigenstates basis. However the interaction with a background
matter is usually given for flavor neutrinos (see, e.g.,
Ref.~\cite{LimMar88}). For the detailed discussion of the explicit
forms of the matrices $(g_{ab}^\mu)$ and $(\mu_{ab})$ in the
flavor eigenstates basis the reader is referred to the recent
review~\cite{Dvo11NOVA}.
%

Since the neutrino mass eigenstates $\psi_a$ are supposed to be
Majorana particles they should obey the Majorana condition in the
form, $\psi_a^c = \mathrm{i} \gamma^2 \psi_a^{*{}} = \varkappa_c
\psi_a$, where $\varkappa_c$ is a phase factor which we shall take
equal to one. If we express the four-component Majorana spinors in
terms of two-component Weyl fields, $\eta_a$ and $\xi_a$, as
\begin{equation}\label{psietaxi}
  \psi_a^{(\eta)} =
  \begin{pmatrix}
    \mathrm{i} \sigma_2 \eta_a^{*{}} \\
    \eta_a \
  \end{pmatrix},
  \quad
  \text{or}
  \quad
  \psi_a^{(\xi)} =
  \begin{pmatrix}
    \xi_a  \\
    - \mathrm{i} \sigma_2 \xi_a^{*{}} \
  \end{pmatrix},
\end{equation}
which satisfy the Majorana condition, we can rewrite
Eq.~\eqref{wepsi} in the two equivalent forms,
\begin{align}\label{weleft}
  \dot{\eta}_a - & (\bm{\sigma}\nabla)\eta_a +
  m_a \sigma_2 \eta_a^{*{}} -
  \mu_{ab} \bm{\sigma}
  (\mathbf{B} - \mathrm{i}\mathbf{E}) \sigma_2 \eta_b^{*{}}
  \notag
  \\
  & +
  \mathrm{i}
  (g^0_{ab}+\bm{\sigma}\mathbf{g}_{ab}) \eta_b = 0,
\end{align}
or
\begin{align}\label{weright}
  \dot{\xi}_a  + & (\bm{\sigma}\nabla)\xi_a -
  m_a \sigma_2 \xi_a^{*{}} +
  \mu_{ab} \bm{\sigma}
  (\mathbf{B} + \mathrm{i}\mathbf{E}) \sigma_2 \xi_b^{*{}}
  \notag
  \\
  & -
  \mathrm{i}
  (g^0_{ab}-\bm{\sigma}^{*{}}\mathbf{g}_{ab}) \xi_b = 0.
\end{align}
Here $\bm{\sigma} = (\sigma_1, \sigma_2, \sigma_3)$ are the Pauli
matrices. In the following we shall postulate these equations.
Note that the analog of Eq.~\eqref{weleft} was previously derived
in Ref.~\cite{DvoMaa09}. We shall choose the spinors $\eta_a$ as
the basic ones since it was experimentally confirmed that active
neutrinos correspond to left-handed particles. Note that the
rigorous proof of the equivalence of Majorana and Weyl fields was
given in Ref.~\cite{FukYan03p289}.

In Ref.~\cite{Dvo11} we demonstrated that the classical dynamics
of a massive Weyl field in vacuum should be described only in
frames of the Hamilton formalism. Generalizing the results of
Ref.~\cite{Dvo11} to include the interaction with a background
matter and an electromagnetic field we arrive to the following
Hamiltonian:
\begin{widetext}
\begin{align}\label{Hamclass}
  H = & \int \mathrm{d}^3\mathbf{r}
  \Big[
  \sum_a
  \big\{
    \pi_a^\mathrm{T} (\bm{\sigma}\nabla) \eta_a -
    (\eta_a^{*{}})^\mathrm{T} (\bm{\sigma}\nabla) \pi_a^{*{}}
    +
    m_a
    \left[
      (\eta_a^{*{}})^\mathrm{T} \sigma_2 \pi_a +
      (\pi_a^{*{}})^\mathrm{T} \sigma_2 \eta_a
    \right]
  \big\}
  \notag
  \displaybreak[2]
  \\
  & +
  \sum_{ab}
  \big\{
    \mu_{ab}
    \big[
      \pi_a^\mathrm{T} \bm{\sigma}
      (\mathbf{B} - \mathrm{i}\mathbf{E}) \sigma_2 \eta_b^{*{}}
      +
      \eta_a^\mathrm{T} \sigma_2 \bm{\sigma}
      (\mathbf{B} + \mathrm{i}\mathbf{E}) \pi_b^{*{}}
    \big]
    -
    \mathrm{i}
    \big[
      \pi_a^\mathrm{T}
      (g^0_{ab}+\bm{\sigma}\mathbf{g}_{ab}) \eta_b
      -
      (\eta_a^{*{}})^\mathrm{T}
      (g^0_{ab}+\bm{\sigma}\mathbf{g}_{ab}) \pi_b^{*{}}
    \big]
  \big\}
  \Big],
\end{align}
where $\pi_a$ are the canonical momenta conjugate to the
``coordinates" $\eta_a$. Using the aforementioned properties of
the matrices $(\mu_{ab})$ and $(g_{ab}^\mu)$ we find that the
functional~\eqref{Hamclass} is real as it should be for a
classical Hamiltonian.
\end{widetext}

Applying the field theory version of the canonical equations to
the Hamiltonian $H$,
\begin{align}
  \label{etaclass}
  \dot{\eta}_a  = & \frac{\delta H}{\delta \pi_a} =
  (\bm{\sigma}\nabla)\eta_a - m_a \sigma_2 \eta_a^{*{}}
  \notag
  \\
  & +
  \mu_{ab} \bm{\sigma}
  (\mathbf{B} - \mathrm{i}\mathbf{E}) \sigma_2 \eta_b^{*{}} -
  \mathrm{i} (g^0_{ab}+\bm{\sigma}\mathbf{g}_{ab}) \eta_b,
  \\
  \label{piclass}
  \dot{\pi}_a  = & - \frac{\delta H}{\delta \eta_a} =
  (\bm{\sigma}^{*{}}\nabla)\pi_a + m_a \sigma_2 \pi_a^{*{}}
  \notag
  \\
  & -
  \mu_{ab} \sigma_2 \bm{\sigma}
  (\mathbf{B} + \mathrm{i}\mathbf{E}) \pi_b^{*{}} +
  \mathrm{i} (g^0_{ab}+\bm{\sigma}^{*{}}\mathbf{g}_{ab}) \pi_b,
\end{align}
one can see that in Eq.~\eqref{etaclass} we reproduce
Eq.~\eqref{weleft} for Weyl particles, which correspond to
left-handed neutrinos, interacting with matter and electromagnetic
field. If we introduce the new variable $\xi_a = \mathrm{i}
\sigma_2 \pi_a$, we can show that Eq.~\eqref{piclass} is
equivalent to Eq.~\eqref{weright} for right-handed neutrinos.

Previous quantum field theory based studies of Majorana neutrinos
in an electromagnetic field and in a background
matter~\cite{SchVal81,Man88} involved the Lagrange formalism. The
mass term in a Lagrangian for the Weyl field $\eta_a$ has the form
(see, e.g., Ref.~\cite{FukYan03p289}),
\begin{equation}
  \mathcal{L}_m = - \frac{\mathrm{i}}{2}
  m_a \eta_a^\mathrm{T} \sigma_2 \eta_a +
  \frac{\mathrm{i}}{2}
  m_a \eta_a^\dag \sigma_2 \eta_a^{*{}}.
\end{equation}
It is however clear that $\mathcal{L}_m$ vanishes if $\eta_a$ is a
first-quantized field having commuting $c$-number components. To
resolve this problem in Refs.~\cite{SchVal81,Man88} it was
supposed that two-component Weyl fields are represented via the
anti-commuting operators. Thus the treatment of Majorana particles
in those works can be considered just as the re-expression of
already quantized fields in terms of the new variables rather than
the generic canonical quantization. Moreover in
Ref.~\cite{SchVal81} it was claimed that ``there is no
`first-quantized' description of a massive two-component field in
terms of $c$-number wave functions".

On the contrary, in the present work we have demonstrated that
classical (first-quantized) massive Weyl fields in presence of an
external electromagnetic field and a background matter can be
perfectly described within the Hamilton formalism, cf.
Eqs.~\eqref{Hamclass}-\eqref{piclass}. This fact just means that
the Lagrange formalism is not a suitable tool for the studies of
Majorana particles. Note that a more detailed description of the
Weyl fields dynamics in vacuum in frames of the canonical approach
as well as the discussion of the applicability of the Lagrange and
the Hamilton formalisms is presented in out recent
work~\cite{Dvo11}.

\section{Quantization\label{QUANT}}

In this section we canonically quantize a Weyl field propagating
in a background matter. On the basis of an exact solution of the
wave equation for a massive Weyl field in matter we express the
energy and the momentum of the field as a sum of contributions of
independent quantum oscillators. From the requirement of the
positive definiteness of the energy it turns out that the
operators in the decomposition of the wave functions should obey
the Fermi-Dirac statistics.

In the following we shall suppose that the background matter in
average is at rest and unpolarized, i.e. $\mathbf{g}_{ab} = 0$.
This approximation is valid in almost all realistic cases. Indeed
the matter motion is relevant for the neutrino dynamics if the
speed of medium is comparable with the speed of light. This
situation may be implemented, e.g., if a beam of neutrinos
propagates inside a relativistic jet from a quasar. Since the
results of the present work are likely to be applied for the
studies of supernova neutrinos, the matter motion seems to be
irrelevant for us.

When neutrinos interact with a non-degenerate plasma, we may
neglect the matter polarization if $\mu_f |\mathbf{B}| \ll T$ (see
Ref.~\cite{NunSemSmiVal97}), where $\mu_f \sim \mu_\mathrm{B}$ is
the magnetic moment of a background fermion, $\mu_\mathrm{B}$ is
the Bohr magneton, and $T$ is the plasma temperature. In the
present work we shall study the influence of both external fields
(see Sec.~\ref{MATTEMF}) and the neutrino self-interaction (see
Sec.~\ref{SI}) on neutrino oscillations. In Ref.~\cite{Dua06} it
was was found that the collective effects in neutrino oscillations
reveal themselves most intensively at the distance $r \sim
100\thinspace\text{km}$ from a protoneutron star. The magnetic
field at this distance can be $B = B_0 (R/r)^3 \sim
10^{10}\thinspace\text{G}$, where $B_0 \sim
10^{13}\thinspace\text{G}$ is the typical magnetic field on the
protoneutron star surface and $R \sim 10\thinspace\text{km}$ is
the stellar radius. Supposing that $T \sim
1\thinspace\text{MeV}$~\cite{Tom03}, we get that the matter
polarization becomes unimportant for stars possessing moderate
magnetic fields.

Let us decompose the Hamiltonian~\eqref{Hamclass} into two terms
$H = H_0 + H_\mathrm{int}$. The former term, $H_0$, contains the
vacuum Hamiltonian as well as the matter interaction term diagonal
in the neutrino types,
\begin{align}\label{Ham0}
  H_0 = & \int \mathrm{d}^3\mathbf{r}
  \sum_a
  \big\{
    \pi_a^\mathrm{T} [(\bm{\sigma}\nabla) - \mathrm{i} g^0_{aa}] \eta_a
    \notag
    \\
    & -
    (\eta_a^{*{}})^\mathrm{T} [(\bm{\sigma}\nabla) - \mathrm{i} g^0_{aa}] \pi_a^{*{}}
    \notag
    \\
    & +
    m_a
    \left[
      (\eta_a^{*{}})^\mathrm{T} \sigma_2 \pi_a +
      (\pi_a^{*{}})^\mathrm{T} \sigma_2 \eta_a
    \right]
  \big\}.
\end{align}
The latter term in this decomposition,
\begin{align}\label{Hamint}
  H_\mathrm{int} = & \int \mathrm{d}^3\mathbf{r}
  \sum_{a \neq b}
  \big\{
    \mu_{ab}
    \big[
      \pi_a^\mathrm{T} \bm{\sigma}
      (\mathbf{B} - \mathrm{i}\mathbf{E}) \sigma_2 \eta_b^{*{}}
      \notag
      \\
      & +
      \eta_a^\mathrm{T} \sigma_2 \bm{\sigma}
      (\mathbf{B} + \mathrm{i}\mathbf{E}) \pi_b^{*{}}
    \big]
    \notag
    \displaybreak[2]
    \\
    & -
    \mathrm{i} g^0_{ab}
    \left[
      \pi_a^\mathrm{T} \eta_b -
      (\eta_a^{*{}})^\mathrm{T} \pi_b^{*{}}
    \right]
  \big\},
\end{align}
has the nondiagonal matter interaction and the interaction with an
electromagnetic field which is nondiagonal by definition.

Analogously to Eqs.~\eqref{etaclass} and~\eqref{piclass} we define
the reduced Hamilton equations which contain only the Hamiltonian
$H_0$: $\dot{\eta}_a^{(0)} = \delta H_0/\delta \pi_a^{(0)}$ and
$\dot{\pi}_a^{(0)} = - \delta H_0/\delta \eta_a^{(0)}$. Using the
results of Refs.~\cite{Dvo11,DvoMaa09} we can find the solutions
of these equations in the form,
\begin{widetext}
\begin{align}\label{etaxisol}
  \eta_a^{(0)}(\mathbf{r},t) = & 
  \int \frac{\mathrm{d}^3\mathbf{p}}{(2\pi)^{3/2}}
  \bigg\{
    \left[
      a_a^{-{}} w_{-{}} e^{-\mathrm{i}E_a^{-{}}t} -
      \frac{m_a}{E_a^{+{}}+|\mathbf{p}|-g^0_{aa}}
      a_a^{+{}} w_{+{}} e^{-\mathrm{i}E_a^{+{}}t}
    \right]
    e^{\mathrm{i}\mathbf{p}\mathbf{r}}
    \notag
    \\
    & +
    \left[
      (a_a^{+{}})^{*{}} w_{-{}} e^{\mathrm{i}E_a^{+{}}t} +
      \frac{m_a}{E_a^{-{}}+|\mathbf{p}|+g^0_{aa}}
      (a_a^{-{}})^{*{}} w_{+{}} e^{\mathrm{i}E_a^{-{}}t}
    \right]
    e^{-\mathrm{i}\mathbf{p}\mathbf{r}}
  \bigg\},
  \notag
  \\
  \xi_a^{(0)}(\mathbf{r},t) = & \mathrm{i} 
  \int \frac{\mathrm{d}^3\mathbf{p}}{(2\pi)^{3/2}}
  \bigg\{
    \left[
      b_a^{+{}} w_{+{}} e^{-\mathrm{i}E_a^{+{}}t} +
      \frac{m_a}{E_a^{-{}}+|\mathbf{p}|+g^0_{aa}}
      b_a^{-{}} w_{-{}} e^{-\mathrm{i}E_a^{-{}}t}
    \right]
    e^{\mathrm{i}\mathbf{p}\mathbf{r}}
    \notag
    \\
    & +
    \left[
      (b_a^{-{}})^{*{}} w_{+{}} e^{\mathrm{i}E_a^{-{}}t} -
      \frac{m_a}{E_a^{+{}}+|\mathbf{p}|-g^0_{aa}}
      (b_a^{+{}})^{*{}} w_{-{}} e^{\mathrm{i}E_a^{+{}}t}
    \right]
    e^{-\mathrm{i}\mathbf{p}\mathbf{r}}
  \bigg\},
\end{align}
where we introduce the new variable $\xi_a^{(0)} = \mathrm{i}
\sigma_2 \pi_a^{(0)}$, $w_{\pm{}}$ are the helicity amplitudes
defined in Ref.~\cite{BerLifPit82}, and
\begin{equation}\label{enWeyl}
  E_a^{(\zeta)} =
  \sqrt{m_a^2 + (|\mathbf{p}| - \zeta g^0_{aa})^2},
\end{equation}
is the energy of a Weyl field~\cite{GriStuTer05,Man88}, $\zeta =
\pm 1$ is the particle helicity. To derive Eqs.~\eqref{etaxisol}
and~\eqref{enWeyl} we suppose that the external field $g^0_{aa}$
is spatially constant.

Using Eq.~\eqref{etaxisol} we can express the
Hamiltonian~\eqref{Ham0} as
\begin{align}\label{Hamquant}
  H_0 = & 
  \int \mathrm{d}^3\mathbf{p}
  \sum_a
  \Big\{
    E_a^{-{}}
    \Big[
      [a_a^{-{}}(\mathbf{p})]^{*{}} b_a^{-{}}(\mathbf{p}) +
      [b_a^{-{}}(\mathbf{p})]^{*{}} a_a^{-{}}(\mathbf{p})
      \notag
      \\
      & +
      \frac{m_a^2}{(E_a^{-{}} + |\mathbf{p}| + g^0_{aa})^2}
      \left\{
        a_a^{-{}}(\mathbf{p}) [b_a^{-{}}(\mathbf{p})]^{*{}} +
        b_a^{-{}}(\mathbf{p}) [a_a^{-{}}(\mathbf{p})]^{*{}}
      \right\}
    \Big]
    \notag
    \\
    & -
    E_a^{+{}}
    \Big[
      a_a^{+{}}(\mathbf{p}) [b_a^{+{}}(\mathbf{p})]^{*{}} +
      b_a^{+{}}(\mathbf{p}) [a_a^{+{}}(\mathbf{p})]^{*{}}
      \notag
      \\
      & +
      \frac{m_a^2}{(E_a^{+{}} + |\mathbf{p}| - g^0_{aa})^2}
      \left\{
        [b_a^{+{}}(\mathbf{p})]^{*{}} a_a^{+{}}(\mathbf{p})  +
        [a_a^{+{}}(\mathbf{p})]^{*{}} b_a^{+{}}(\mathbf{p})
      \right\}
    \Big]
    \notag
    \displaybreak[2]
    \\
    & +
    \mathrm{i} m_a
    \Big\{
      \frac{E_a^{-{}}}{E_a^{-{}} + |\mathbf{p}| + g^0_{aa}}
      \big[
        e^{-2\mathrm{i}E_a^{-{}}t}
        \left\{
          a_a^{-{}}(\mathbf{p}) b_a^{-{}}(-\mathbf{p}) +
          b_a^{-{}}(-\mathbf{p}) a_a^{-{}}(\mathbf{p})
        \right\}
        \notag
        \\
        & +
        e^{2\mathrm{i}E_a^{-{}}t}
        \left\{
          [a_a^{-{}}(\mathbf{p})]^{*{}} [b_a^{-{}}(-\mathbf{p})]^{*{}} +
          [b_a^{-{}}(-\mathbf{p})]^{*{}} [a_a^{-{}}(\mathbf{p})]^{*{}}
        \right\}
      \big]
      \notag
      \\
      & +
      \frac{E_a^{+{}}}{E_a^{+{}} + |\mathbf{p}| - g^0_{aa}}
      \big[
        e^{-2\mathrm{i}E_a^{+{}}t}
        \left\{
          a_a^{+{}}(\mathbf{p}) b_a^{+{}}(-\mathbf{p}) +
          b_a^{+{}}(-\mathbf{p}) a_a^{+{}}(\mathbf{p})
        \right\}
        \notag
        \\
        &
        +
        e^{2\mathrm{i}E_a^{+{}}t}
        \left\{
          [a_a^{+{}}(\mathbf{p})]^{*{}} [b_a^{+{}}(-\mathbf{p})]^{*{}} +
          [b_a^{+{}}(-\mathbf{p})]^{*{}} [a_a^{+{}}(\mathbf{p})]^{*{}}
        \right\}
      \big]
    \Big\}
  \Big\},
\end{align}
\end{widetext}
in terms of the creation, $[a_a^{\pm{}}(\mathbf{p})]^{*{}}$ and
$[b_a^{\pm{}}(\mathbf{p})]^{*{}}$, as well as the annihilation,
$a_a^{\pm{}}(\mathbf{p})$ and $b_a^{\pm{}}(\mathbf{p})$,
operators. Note that the operators $a_a^{\pm{}}(\mathbf{p})$ and
$b_a^{\pm{}}(\mathbf{p})$ are independent up to now.

Let us establish the following relation:
\begin{equation}\label{Majcondquant}
  a_a^{\pm{}}(\mathbf{p})(E_a^{\pm{}} + |\mathbf{p}| \mp g^0_{aa}) =
  4 b_a^{\pm{}}(\mathbf{p})(|\mathbf{p}| \mp g^0_{aa}),
\end{equation}
and the analogous expression for conjugate operators, as well as
suggest that the operators $a_a^{\pm{}}(\mathbf{p})$ obey the
anti-commutation properties,
\begin{equation}\label{anticomm}
  \{
    a_a^{\pm{}}(\mathbf{k}); [a_b^{\pm{}}(\mathbf{p})]^{*{}}
  \}_{+{}} =
  \delta_{ab} \delta^3(\mathbf{k} - \mathbf{p}),
\end{equation}
with all the rest of the anticommutators being equal to zero. In
this case the time dependent terms in Eq.~\eqref{Hamquant} are
washed out. The remaining terms can be represented as
\begin{align}\label{totenquant}
  H_0 = & \int \mathrm{d}^3\mathbf{p}
  \sum_a
  [E_a^{-{}} (a_a^{-{}})^{*{}} a_a^{-{}} +
  E_a^{+{}} (a_a^{+{}})^{*{}} a_a^{+{}}]
  \notag
  \\
  & + \text{divergent terms},
\end{align}
which shows that the total energy of a massive Weyl field is a sum
of energies corresponding to elementary oscillators of positive
and negative helicities.

Using the results of Ref.~\cite{Dvo11} we can also quantize the
total momentum of a Weyl field defined as
\begin{align}\label{momdef}
  \mathbf{P}_0 = & \int \mathrm{d}^3\mathbf{r}
  \sum_a
  \bigg[
    \left( \eta_a^{(0)*{}} \right)^\mathrm{T} \nabla \pi_a^{(0)*{}}
    \notag
    \\
    & -
    \left( \pi_a^{(0)} \right)^\mathrm{T} \nabla \eta_a^{(0)}
  \bigg].
\end{align}
With help of Eqs.~\eqref{etaxisol}, \eqref{Majcondquant},
and~\eqref{anticomm} we rewrite Eq.~\eqref{momdef} in the
following form:
\begin{align}\label{totmomquant}
  \mathbf{P}_0 = & \int \mathrm{d}^3\mathbf{p}
  \sum_a
  \mathbf{p}
  [(a_a^{-{}})^{*{}} a_a^{-{}} +
  (a_a^{+{}})^{*{}} a_a^{+{}}]
  \notag
  \\
  & + \text{divergent terms},
\end{align}
which has the similar structure as Eq.~\eqref{totenquant}.

Note that the divergent terms in Eqs.~\eqref{totenquant}
and~\eqref{totmomquant} contain the factor $\delta^3(\mathbf{p} =
0) \to \infty$, which can be formally removed by the normal
ordering of the operators $a_a^{\pm{}}$ and $(a_a^{\pm{}})^{*{}}$.
It is also interesting to mention that a massive Weyl field in
vacuum can be quantized in the two independent ways (see
Ref.~\cite{Dvo11}) because of the degeneracy of the neutrino
energy levels: $E_a^{-{}} = E_a^{+{}} = \sqrt{m_a^2 +
|\mathbf{p}|^2}$. On the contrary, in matter only one of the
possibilities for the quantization gives the correct result for
the total energy~\eqref{totenquant} since the energy levels are no
longer degenerate, cf. Eq.~\eqref{enWeyl}.

\section{Nondiagonal interaction with matter and electromagnetic field\label{MATTEMF}}

In this section we apply the approach for the quantization of a
massive Weyl in a background matter developed in Sec.~\ref{QUANT}
for the treatment of the nondiagonal Hamiltonian $H_\mathrm{int}$
given in Eq.~\eqref{Hamint}. On the basis of the obtained results
and using the density matrix formalism~\cite{SigRaf93} we derive
the effective Hamiltonian for the description of neutrino
spin-flavor oscillations in matter and electromagnetic field. Then
we demonstrate that for ultrarelativistic particles our effective
Hamiltonian is consistent with the analogous expression obtained
in frames of the standard quantum mechanical approach.

To quantize the Hamiltonian $H_\mathrm{int}$ we shall use the
forward scattering approximation. It means that one has to account
for only the terms conserving the number of
particles~\cite{Raf96p318}. Using  Eqs.~\eqref{etaxisol},
\eqref{Majcondquant}, and~\eqref{anticomm} we rewrite
Eq.~\eqref{Hamint} in the form,
\begin{widetext}
\begin{equation}\label{Hamintquant}
  H_\mathrm{int} = \int \mathrm{d}^3\mathbf{p}
  \sum_{a \neq b}
  [M_{ab}^{-{}} (a_a^{-{}})^{*{}} a_b^{-{}}
  e^{\mathrm{i}\delta_{ab}^{-}t} +
  M_{ab}^{+{}} (a_a^{+{}})^{*{}} a_b^{+{}}
  e^{\mathrm{i}\delta_{ab}^{+}t}
  +
  F_{ab}^{-{}} (a_a^{-{}})^{*{}} a_b^{+{}}
  e^{\mathrm{i}\sigma_{ab}^{-}t} +
  F_{ab}^{+{}} (a_a^{+{}})^{*{}} a_b^{-{}}
  e^{\mathrm{i}\sigma_{ab}^{+}t}],
\end{equation}
where $\delta_{ab}^{\pm{}} = E_a^{\pm{}} - E_b^{\pm{}}$,
$\sigma_{ab}^{\pm{}} = E_a^{\pm{}} - E_b^{\mp{}}$,
\begin{align}\label{Mab}
  M_{ab}^{\pm{}} = & \mp \frac{g_{ab}}{4}
  \left(
    \frac{E_a^{\pm{}} + |\mathbf{p}| \mp g^0_{aa}}{|\mathbf{p}| \mp g^0_{aa}} +
    \frac{E_b^{\pm{}} + |\mathbf{p}| \mp g^0_{bb}}{|\mathbf{p}| \mp g^0_{bb}}
  \right)
  \left[
    1 + \frac{m_a m_b}
    {(E_a^{\pm{}} + |\mathbf{p}| \mp g^0_{aa})
    (E_b^{\pm{}} + |\mathbf{p}| \mp g^0_{bb})}
  \right]
  \notag
  \displaybreak[2]
  \\
  & \pm
  \frac{\mu_{ab}}{4}
  \bigg\{
  \frac{1}{|\mathbf{p}| \mp g^0_{aa}}
  \bigg[
    m_a w_{+{}}^\mathrm{T}
    (\bm{\sigma}[\mathbf{B} \mp \mathrm{i}\mathbf{E}])
    w_{+{}}^{*{}}
    +
    m_b w_{-{}}^\mathrm{T}
    (\bm{\sigma}[\mathbf{B} \pm \mathrm{i}\mathbf{E}])
    w_{-{}}^{*{}}
    \frac{E_a^{\pm{}} + |\mathbf{p}| \mp g^0_{aa}}
    {E_b^{\pm{}} + |\mathbf{p}| \mp g^0_{bb}}
  \bigg]
  \notag
  \displaybreak[2]
  \\
  & +
  \frac{1}{|\mathbf{p}| \mp g^0_{bb}}
  \bigg[
    m_b w_{+{}}^\mathrm{T}
    (\bm{\sigma}[\mathbf{B} \pm \mathrm{i}\mathbf{E}])
    w_{+{}}^{*{}}
    +
    m_a w_{-{}}^\mathrm{T}
    (\bm{\sigma}[\mathbf{B} \mp \mathrm{i}\mathbf{E}])
    w_{-{}}^{*{}}
    \frac{E_b^{\pm{}} + |\mathbf{p}| \mp g^0_{bb}}
    {E_a^{\pm{}} + |\mathbf{p}| \mp g^0_{aa}}
  \bigg]
  \bigg\},
\end{align}
and
\begin{align}\label{Fab}
  F_{ab}^{\pm{}} = & \frac{\mu_{ab}}{4}
  \bigg\{
    w_{\pm{}}^\mathrm{T}
    (\bm{\sigma}[\mathbf{B} \pm \mathrm{i}\mathbf{E}])
    w_{\mp{}}^{*{}}
    \left[
      \frac{E_a^{\pm{}} + |\mathbf{p}| \mp g^0_{aa}}
      {|\mathbf{p}| \mp g^0_{aa}} +
      \frac{E_b^{\mp{}} + |\mathbf{p}| \pm g^0_{bb}}
      {|\mathbf{p}| \pm g^0_{bb}}
    \right]
      \notag
      \\
      & -
      m_a m_b
      w_{\mp{}}^\mathrm{T}
      (\bm{\sigma}[\mathbf{B} \mp \mathrm{i}\mathbf{E}])
      w_{\pm{}}^{*{}}
    \bigg[
      \frac{1}{(E_a^{\pm{}} + |\mathbf{p}| \mp g^0_{aa})
      (|\mathbf{p}| \pm g^0_{bb})}
      +
      \frac{1}{(E_b^{\mp{}} + |\mathbf{p}| \pm g^0_{bb})
      (|\mathbf{p}| \mp g^0_{aa})}
    \bigg]
  \bigg\}.
\end{align}
\end{widetext}
Note that Eqs.~\eqref{Hamintquant}-\eqref{Fab} are valid for
arbitrary neutrino masses, momentum, and the diagonal neutrino
interaction with matter.

Now we define the neutrino density matrix as
\begin{equation}\label{rhodef}
  \delta^3(\mathbf{p}-\mathbf{k}) \rho_{AB}(\mathbf{p}) =
  \langle a^{*{}}_B(\mathbf{p}) a_A(\mathbf{k}) \rangle,
\end{equation}
where $A = (\zeta, a)$ is a composite index and $\langle \dots
\rangle$ is the statistical averaging over the neutrino ensemble.
In principle, we could interchange the indexes $A$ and $B$ in the
rhs of Eq.~\eqref{rhodef}. Since we study the dynamics of Majorana
neutrinos, such a transposition would mean the consideration of
neutrinos as antiparticles rather than as particles as we do here.
Anyway both definitions will give equivalent results. Note that a
density matrix, having both neutrino type and helicity indexes,
which is analogous to Eq.~\eqref{rhodef}, was studied in
Ref.~\cite{Saw11} where the interaction between Dirac neutrinos,
mediated by a scalar boson, was discussed.

Applying the quantum Liouville equation for the description of the
density matrix evolution,
\begin{equation}\label{qLiouv}
  \mathrm{i} \dot{\rho} = [\rho, H_\mathrm{int}],
\end{equation}
we can rewrite it as $\mathrm{i} \dot{\rho} = [\mathcal{H}, \rho]$
using the effective quantum mechanical Hamiltonian,
\begin{equation}\label{Hinteff}
  \mathcal{H} =
  \begin{pmatrix}
    (M_{ab}^{-{}} e^{\mathrm{i}\delta_{ab}^{-}t}) &
    (F_{ab}^{-{}} e^{\mathrm{i}\sigma_{ab}^{-}t}) \\
    (F_{ab}^{+{}} e^{\mathrm{i}\sigma_{ab}^{+}t}) &
    (M_{ab}^{+{}} e^{\mathrm{i}\delta_{ab}^{+}t}) \
  \end{pmatrix}.
\end{equation}
Again we stress that yet no expansion over the parameter
$m_a/|\mathbf{p}|$, which is small for ultrarelativistic
neutrinos, is made. Thus Eqs.~\eqref{qLiouv} and~\eqref{Hinteff}
are valid for the description of neutrinos with arbitrary initial
momentum.

To study the evolution of our system we use Eq.~\eqref{etaxisol},
where the wave functions already contain time dependent
exponential factors. The energies in Eq.~\eqref{etaxisol}
correspond to the total diagonal Hamiltonian~\eqref{Ham0}, cf.
Eqs.~\eqref{enWeyl} and~\eqref{totenquant}, which contains both
the mass term and the diagonal interaction with a background
matter rather than only a kinetic term as in Ref.~\cite{SigRaf93}.
Thus our treatment is analogous to the Dirac picture of the
quantum theory. That is why in Eq.~\eqref{qLiouv} it is sufficient
to commute the density matrix only with $H_\mathrm{int}$ rather
than with the total Hamiltonian $H = H_0 + H_\mathrm{int}$.

To derive Eqs.~\eqref{Hamintquant}-\eqref{Fab} we suppose that the
external fields $g^0_{ab}$, $a \neq b$, and $F_{\mu\nu}$ are
spatially constant. In fact, for this supposition to be valid, the
characteristic scale of the external field variation
$L_\mathrm{ext}$ should be much bigger than the typical width of
the neutrino wave packet $\hbar/E_\nu$. One can see that in almost
all realistic situations external fields can be regarded as
constant for the derivation of an effective Hamiltonian.
Nevertheless, when the obtained effective Hamiltonian is used to
describe the dynamics of the system, e.g., with help of
Eq.~\eqref{qLiouv}, we should account for the variation of
external fields.

To demonstrate the consistency of the obtained results with the
achievements of the standard quantum mechanical description of
neutrino spin-flavor oscillations (see, e.g.,
Ref.~\cite{LimMar88}) we discuss the simplest case of the two
neutrino eigenstates, $a=1,2$, and consider the situation of
ultrarelativistic particles, $|\mathbf{k}| \gg \max(m_{a},
g^0_{aa})$, where $\mathbf{k}$ is the initial momentum of
neutrinos. In this case we should decompose the energy
levels~\eqref{enWeyl} as
\begin{equation}
  E_a^{\pm{}} = |\mathbf{k}| + \frac{m_a^2}{2|\mathbf{k}|} \mp
  g^0_{aa} + \dotsb .
\end{equation}
Then, supposing that $\mathbf{E} = 0$, since it is difficult to
create a large scale electric field, we get that $ M_{ab}^{\pm{}}
\approx \mp g_{ab}$ and $F_{ab}^{\pm{}} \approx - \mu_{ab}
|\mathbf{B}| \sin \vartheta_\mathbf{kB}$, where
$\vartheta_\mathbf{kB}$ is the angle between the vectors
$\mathbf{k}$ and $\mathbf{B}$.

To eliminate the time dependent factors in the effective
Hamiltonian~\eqref{Hinteff} we make the transformation of the
density matrix~\cite{DvoMaa09},
\begin{widetext}
\begin{equation}\label{matrtransf}
  \rho_\mathrm{qm} =
  \mathcal{U} \rho \mathcal{U}^\dag,
  \quad
  \mathcal{U} = \mathrm{diag}\{e^{-\mathrm{i}(\Phi+g^0_{11})t},
  e^{\mathrm{i}(\Phi-g^0_{22})t},
  e^{-\mathrm{i}(\Phi-g^0_{11})t},
  e^{\mathrm{i}(\Phi+g^0_{22})t}\},
\end{equation}
where $\Phi = \delta m^2 / 4|\mathbf{k}|$ is the phase of vacuum
oscillations and $\delta m^2 = m_1^2 -m_2^2$ is the mass squared
difference.

The evolution of the transformed density matrix can be represented
as $\mathrm{i} \dot{\rho}_\mathrm{qm} = [\mathcal{H}_\mathrm{qm},
\rho_\mathrm{qm}]$, where the new effective Hamiltonian has the
form,
\begin{equation}\label{Heffqm}
  \mathcal{H}_\mathrm{qm} =
  \mathcal{U} \mathcal{H} \mathcal{U}^\dag +
  \mathrm{i} \dot{\mathcal{U}}\mathcal{U}^\dag
  =
  \begin{pmatrix}
    \Phi + g^0_{11} & g^0_{12} &
    0 & -\mu_{12} |\mathbf{B}| \sin \vartheta_\mathbf{kB} \\
    g^0_{21} & - \Phi + g^0_{22} &
    -\mu_{21} |\mathbf{B}| \sin \vartheta_\mathbf{kB} & 0 \\
    0 & -\mu_{12} |\mathbf{B}| \sin \vartheta_\mathbf{kB} &
    \Phi - g^0_{11} & - g^0_{12} \\
    -\mu_{21} |\mathbf{B}| \sin \vartheta_\mathbf{kB} & 0 &
    - g^0_{21} & - \Phi - g^0_{22} \
  \end{pmatrix}.
\end{equation}
\end{widetext}
Recalling the properties of the magnetic moments matrix for
Majorana neutrinos, $\mu_{12} = \mathrm{i}\mu$ and $\mu_{21} = -
\mathrm{i}\mu$, with $\mu$ being a real number, we can see that
Eq.~\eqref{Heffqm} reproduces the well known quantum mechanical
Hamiltonian for spin-flavor oscillations of Majorana neutrinos in
matter and a magnetic field~\cite{LimMar88}.

Note that previously calculated transition probabilities of
neutrino oscillations in a magnetic field~\cite{SchVal81} and in a
background matter~\cite{Man88} can be re-derived using the
effective Hamiltonian~\eqref{Heffqm} which was obtained in frames
of the approach involving canonically quantized Weyl fields.
However, as we mentioned in Sec.~\ref{CLASS}, methodologically our
method for the description of massive Majorana neutrinos in
external fields is more logical since it is based on the first
principles of the quantum field theory.

\section{Self-interaction\label{SI}}

In this section we generalize the results of
Secs.~\ref{CLASS}-\ref{MATTEMF} to include the neutrino
self-interaction. First we re-formulate the previously proposed
Hamiltonian for the self-interaction in terms of the two-component
Weyl fields and then we quantize it. Finally, using the density
matrix formalism we derive the corresponding contribution to the
quantum mechanical effective Hamiltonian and compare it with the
previously obtained results.

The Hamiltonian describing the neutrino self-interaction, mediated
by a neutral $Z$-boson, was derived in Refs.~\cite{Sam93,SigRaf93}
and has the form,
\begin{equation}\label{HSpsi}
  H_\mathrm{S} = \int \mathrm{d}^3\mathbf{r}
  \sum_{abcd}
  G_{ab} G_{cd}
  \bar{\psi}_a \gamma_\mu^\mathrm{L} \psi_b
  \cdot
  \bar{\psi}_c \gamma^\mu_\mathrm{L} \psi_d,
\end{equation}
where $\gamma_\mu^\mathrm{L} = \gamma_\mu (1 - \gamma^5)/2$ and
$G_{ab}$ are the coefficients which depend on the neutrino
interactions channel. The explicit form of these coefficients can
be found in Ref.~\cite{Raf96p583}. Let us choose the wave function
$\psi_a$ in Eq.~\eqref{HSpsi} as $\psi_a^{(\eta)}$ in
Eq.~\eqref{psietaxi}. Therefore we express the
Hamiltonian~\eqref{HSpsi} in terms of the two-component Weyl
spinors as
\begin{widetext}
\begin{equation}\label{HSeta}
  H_\mathrm{S} = \int \mathrm{d}^3\mathbf{r}
  \sum_{abcd}
  G_{ab} G_{cd}
  \eta_a^\dag \sigma_\mu \eta_b
  \cdot
  \eta_c^\dag \sigma^\mu \eta_d,
\end{equation}
where $\sigma_\mu = (1,\bm{\sigma})$. It should be noted that we
presented the heuristic derivation of Eq.~\eqref{HSeta} from
Eq.~\eqref{HSpsi}. In principle we could just postulate
Eq.~\eqref{HSeta}.

Using Eqs.~\eqref{etaxisol} and~\eqref{Majcondquant} we cast the
self-interaction Hamiltonian~\eqref{HSeta} into the form,
\begin{equation}
  H_\mathrm{S} = \frac{1}{(2\pi)^3} \int
  \mathrm{d}^3\mathbf{p} \mathrm{d}^3\mathbf{p}'
  \mathrm{d}^3\mathbf{q} \mathrm{d}^3\mathbf{q}'
  \delta(\mathbf{p} + \mathbf{q} - \mathbf{p}' - \mathbf{q}')
  \sum_{abcd}
  G_{ab} G_{cd}
  [\eta_a(\mathbf{q})^{*{}}]^\mathrm{T} \sigma_\mu
  \eta_b(\mathbf{q}')
  \cdot
  [\eta_c(\mathbf{p})^{*{}}]^\mathrm{T} \sigma^\mu
  \eta_d(\mathbf{p}'),
\end{equation}
where
\begin{equation}
  \eta_a(\mathbf{p}) = \sum_{\zeta = \pm 1}
  \Big[
    a_a^{(\zeta)}(\mathbf{p}) u_a^{(\zeta)}(\mathbf{p})
    e^{-\mathrm{i}E_a^{(\zeta)}t}
    +
    \left[ a_a^{(\zeta)}(-\mathbf{p}) \right]^{*{}}
    v_a^{(\zeta)}(-\mathbf{p}) e^{\mathrm{i}E_a^{(\zeta)}t}
  \Big],
\end{equation}
is the Fourier transform of the wave function $\eta_a$ and
\begin{equation}
  u_a^{-{}}(\mathbf{p}) = w_{-{}}(\mathbf{p}),
  \quad
  v_a^{+{}}(\mathbf{p}) = w_{-{}}(\mathbf{p}),
  \quad
  u_a^{+{}}(\mathbf{p}) =
  - \frac{m_a}{E_a^{+{}} + |\mathbf{p}| - g_{aa}^0}
  w_{+{}}(\mathbf{p}),
  \quad
  v_a^{-{}}(\mathbf{p}) =
  \frac{m_a}{E_a^{+{}} + |\mathbf{p}| + g_{aa}^0}
  w_{+{}}(\mathbf{p}),
\end{equation}
are the basis spinors rewritten in a formalized manner, cf.
Eq.~\eqref{etaxisol}.

Applying Eq.~\eqref{qLiouv} to account for the contribution of the
self-interaction to the dynamics of the system and again working
in the forward scattering limit~\cite{Raf96p318}, we can represent
the evolution of the density matrix, $\mathrm{i} \dot{\rho} =
[\mathcal{H}_\mathrm{S}, \rho]$, using the effective Hamiltonian
$\mathcal{H}_\mathrm{S}$. After a bit lengthy but straightforward
calculations we get the following expression for
$\mathcal{H}_\mathrm{S}$:
\begin{align}\label{HSquant}
  \mathcal{H}_\mathrm{S} = & 2
  \int \frac{\mathrm{d}^3\mathbf{p}}{(2\pi)^3}
  \big\{
    M_\mu(\mathbf{k},\mathbf{k})
    \mathrm{tr} \{[M^\mu(\mathbf{p},\mathbf{p}) -
    N^\mu(\mathbf{p},\mathbf{p})] \rho(\mathbf{p})\}
    +
    N_\mu(\mathbf{k},\mathbf{k})
    \mathrm{tr} \{[N^\mu(\mathbf{p},\mathbf{p}) -
    M^\mu(\mathbf{p},\mathbf{p})] \rho(\mathbf{p})\}
    \\
    \notag
    & -
    [M_\mu(\mathbf{k},\mathbf{p}) -
    N_\mu(\mathbf{k},\mathbf{p})]
    \rho(\mathbf{p})
    [M^\mu(\mathbf{p},\mathbf{k}) -
    N^\mu(\mathbf{p},\mathbf{k})]
    +
    (K_\mu(\mathbf{k},\mathbf{p}) -
    [K_\mu(\mathbf{k},\mathbf{p})]^\mathrm{T})
    \rho^\mathrm{T}(\mathbf{p})
    (L^\mu(\mathbf{p},\mathbf{k}) -
    [L^\mu(\mathbf{p},\mathbf{k})]^\mathrm{T})
  \big\},
\end{align}
where
\begin{align}\label{MNKL}
  M_{AB}^\mu(\mathbf{p},\mathbf{k}) = & G_{ab}
  e^{\mathrm{i}E_A(\mathbf{p})t}
  \langle u_A(\mathbf{p}) | \sigma^\mu | u_B(\mathbf{k}) \rangle
  e^{-\mathrm{i}E_B(\mathbf{k})t},
  \quad
  K_{AB}^\mu(\mathbf{p},\mathbf{k}) = G_{ab}
  e^{\mathrm{i}E_A(\mathbf{p})t}
  \langle u_A(\mathbf{p}) | \sigma^\mu | v_B(\mathbf{k}) \rangle
  e^{\mathrm{i}E_B(\mathbf{k})t},
  \notag
  \\
  N_{AB}^\mu(\mathbf{p},\mathbf{k}) = & G_{ba}
  e^{-\mathrm{i}E_B(\mathbf{k})t}
  \langle v_B(\mathbf{k}) | \sigma^\mu | u_A(\mathbf{p}) \rangle
  e^{\mathrm{i}E_A(\mathbf{p})t},
  \quad
  L_{AB}^\mu(\mathbf{p},\mathbf{k}) = G_{ab}
  e^{-\mathrm{i}E_A(\mathbf{p})t}
  \langle v_A(\mathbf{p}) | \sigma^\mu | u_B(\mathbf{k}) \rangle
  e^{-\mathrm{i}E_B(\mathbf{k})t}.
\end{align}
In Eq.~\eqref{HSquant} the transposition means the interchange of
both discrete and continuous indexes, i.e.
$[L_{AB}^\mu(\mathbf{p},\mathbf{k})]^\mathrm{T} = G_{ba}
e^{-\mathrm{i}E_B(\mathbf{k})t} \langle v_B(\mathbf{k}) |
\sigma^\mu | u_A(\mathbf{p}) \rangle
e^{-\mathrm{i}E_A(\mathbf{p})t}$ etc.

Note that Eqs.~\eqref{HSquant} and~\eqref{MNKL} are valid for
arbitrary neutrino masses, initial momenta, and the diagonal
neutrino interaction with matter. However the analysis of these
expressions is quite cumbersome. That is why again we discuss the
situation of the two neutrino generations, $a = 1, 2$, and suppose
that neutrinos are ultrarelativistic particles, $|\mathbf{k}| \gg
\max(m_{a}, g^0_{aa})$. Then, to eliminate the time dependence in
Eq.~\eqref{MNKL} we make the additional matrix transformation of
the effective Hamiltonian, $\mathcal{H}_\mathrm{qm} = \mathcal{U}
\mathcal{H}_\mathrm{S} \mathcal{U}^\dag$, where
$\mathcal{H}_\mathrm{S} = \mathcal{H}_\mathrm{S}[\mathcal{U} \rho
\mathcal{U}^\dag]$, since $\mathcal{H}_\mathrm{S}$ is the function
of $\rho$, and the matrix $\mathcal{U}$ is defined in
Eq.~\eqref{matrtransf}.

Finally we can represent the contribution of the neutrino
self-interaction to the quantum mechanical effective Hamiltonian
as,
\begin{equation}\label{HSqm}
  \mathcal{H}_\mathrm{qm}(\mathbf{k}) =
  \mathrm{diag}(\mathcal{H}_{-{}-{}}, \mathcal{H}_{+{}+{}}),
\end{equation}
where
\begin{align}\label{HSqmcomp}
  \mathcal{H}_{-{}-{}} = & 2
  \int \frac{\mathrm{d}^3\mathbf{p}}{(2\pi)^3}
  (1-\cos\vartheta_\mathbf{kp})
  \big\{
    G \mathrm{tr}[G \rho_{-{}-{}}(\mathbf{p}) -
    G^\mathrm{T} \rho_{+{}+{}}(\mathbf{p})]
    +
    G [\rho_{-{}-{}}(\mathbf{p}) -
    \rho_{+{}+{}}^\mathrm{T}(\mathbf{p})] G
  \big\},
  \notag
  \displaybreak[2]
  \\
  \mathcal{H}_{+{}+{}} = & 2
  \int \frac{\mathrm{d}^3\mathbf{p}}{(2\pi)^3}
  (1-\cos\vartheta_\mathbf{kp})
  \big\{
    G^\mathrm{T} \mathrm{tr}[G^\mathrm{T} \rho_{+{}+{}}(\mathbf{p}) -
    G \rho_{-{}-{}}(\mathbf{p})]
    +
    G^\mathrm{T} [\rho_{+{}+{}}(\mathbf{p}) -
    \rho_{-{}-{}}^\mathrm{T}(\mathbf{p})] G^\mathrm{T}
  \big\},
\end{align}
and $\mathcal{H}_{\pm{}\mp{}} = 0$. Here $\vartheta_\mathbf{kp}$
is the angle between the vectors $\mathbf{k}$ and $\mathbf{p}$ and
we use the helicity components of the density matrix,
\end{widetext}
\begin{equation}
  \rho_\mathrm{qm} =
  \begin{pmatrix}
    \rho_{-{}-{}} & \rho_{-{}+{}} \\
    \rho_{+{}-{}} & \rho_{+{}+{}} \
  \end{pmatrix}.
\end{equation}
To derive Eqs.~\eqref{HSqm} and~\eqref{HSqmcomp} we use the
identity for two-component $c$-number spinors, $\eta_1^\dag
\sigma_\mu \eta_2 \cdot \eta_3^\dag \sigma^\mu \eta_4 = -
\eta_1^\dag \sigma_\mu \eta_4 \cdot \eta_2^\dag \sigma^\mu
\eta_3$, which results from the Fierz transformation of
four-component spinors, $\bar{\psi}_1 \gamma_\mu (1 - \gamma^5)
\psi_2 \cdot \bar{\psi}_3 \gamma^\mu (1 - \gamma^5) \psi_4 = -
\bar{\psi}_1 \gamma_\mu (1 - \gamma^5) \psi_4 \cdot \bar{\psi}_2
\gamma^\mu (1 - \gamma^5) \psi_3$.

One can conclude from Eq.~\eqref{HSqmcomp} that the
self-interaction influences spin-flavor oscillations of neutrinos.
However, since the nondiagonal terms in Eq.~\eqref{HSqm} are equal
to zero for ultrarelativistic particles, the self-interaction
cannot directly induce transitions between different helicity
states.

It should be noted that in the majority of works where collective
effects in neutrino oscillations were studied, the case of Dirac
neutrinos was examined. Although one can expect that in the
ultrarelativistic case the dynamics of Dirac and Majorana
neutrinos should be similar, we cannot reach a complete
coincidence because Dirac particles have twice more degrees of
freedom, i.e. an additional density matrix for antineutrinos is
required. Nevertheless let us check the consistency of our
findings with the previously obtained results. First we should
chose a definite helicity. For example, we may put
$\mathcal{H}_{-{}-{}} \neq 0$ and $\mathcal{H}_{+{}+{}} = 0$.
Then, defining the ``antineutrino" density matrix as $\bar{\rho} =
\rho_{+{}+{}}^\mathrm{T}$, cf. Eq.~\eqref{rhodef}, we re-derive
the contribution of the neutrino self-interaction to the effective
Hamiltonian obtained in Ref.~\cite{SigRaf93}.

At the end of this section we notice that the presented derivation
of Eqs.~\eqref{HSqm} and~\eqref{HSqmcomp} is not unique. In the
general self-interaction Hamiltonian~\eqref{HSpsi} we can set
$\psi_a = \psi_a^{(\xi)}$, where $\psi_a^{(\xi)}$ is defined in
Eq.~\eqref{psietaxi}. Thus the Hamiltonian $H_\mathrm{S}$ can be
expressed in terms of the canonical momenta,
\begin{equation}\label{HSpi}
  H_\mathrm{S} = \int \mathrm{d}^3\mathbf{r}
  \sum_{abcd}
  G_{ab} G_{cd}
  \pi_a^\mathrm{T} \sigma_\mu \pi_b^{*{}}
  \cdot
  \pi_c^\mathrm{T} \sigma^\mu \pi_d^{*{}},
\end{equation}
where use the relation between $\xi_a$ and $\pi_a$: $\xi_a =
\mathrm{i} \sigma_2 \pi_a$ (see also Sec.~\ref{CLASS}).

Then we should re-define the wave functions $\eta_a$ and $\xi_a$
in Eq.~\eqref{etaxisol}, introducing the additional multiplier
$1/2$ in each spinor, as well as the connection between operators
$a_a^{\pm{}}(\mathbf{p})$ and $b_a^{\pm{}}(\mathbf{p})$, which now
reads, $a_a^{\pm{}}(\mathbf{p})(E_a^{\pm{}} + |\mathbf{p}| \mp
g^0_{aa}) = b_a^{\pm{}}(\mathbf{p})(|\mathbf{p}| \mp g^0_{aa})$,
cf. Eq.~\eqref{Majcondquant}. Note that these modifications do not
affect the results of Secs.~\ref{QUANT} and~\ref{MATTEMF}.
Performing the same calculations which led us to Eqs.~\eqref{HSqm}
and~\eqref{HSqmcomp}, but using the modified
Hamiltonian~\eqref{HSpi}, we get that for ultrarelativistic
particles the contribution of the neutrino self-interaction to the
quantum mechanical effective Hamiltonian coincides with
Eqs.~\eqref{HSqm} and~\eqref{HSqmcomp}.

\section{Conclusion\label{CONCL}}

In summary me mention that in the present work we have constructed
the consistent quantum theory of a system of massive Weyl fields
propagating in a background matter and interacting between
themselves and with an external electromagnetic field. We have
obtained several important results.

First, in Sec.~\ref{CLASS}, the classical field theory description
of a massive Weyl field in an arbitrarily moving and polarized
matter and an electromagnetic field has been presented. Using the
approach developed in Ref.~\cite{Dvo11}, where the evolution of a
massive Weyl field in vacuum was studied, we have derived the
classical Hamiltonian~\eqref{Hamclass} for our system. Then
applying the canonical Hamilton equations~\eqref{etaclass}
and~\eqref{piclass} we have re-obtained the analog of the well
known wave equation~\eqref{wepsi} for a Majorana neutrino in
matter and an electromagnetic field. This our result corrects the
previous statement~\cite{SchVal81} that massive Majorana particles
are essentially quantum objects described only using the creation
and annihilation operators formalism. Moreover, now we have
expanded the classical field theory approach, cf.
Ref.~\cite{Dvo11}, to include the interaction with matter and an
electromagnetic field.

Second, in Sec.~\ref{QUANT}, we have canonically quantized massive
Weyl fields in a non-moving and unpolarized matter. We have used
the plane wave solution~\eqref{etaxisol} and~\eqref{enWeyl} (see
also Ref.~\cite{DvoMaa09}) of the corresponding wave equation,
where we supposed that the expansion coefficients are the
operators. Then, requiring the positive definiteness of the total
energy~\eqref{Hamquant}, we have obtained that the operator
expansion coefficients should satisfy the anticommutation
properties~\eqref{anticomm}. It is interesting to mention that
unlike the quantization of a Weyl field in vacuum, where two
independent quantization schemes are possible, in matter there is
only one opportunity~\eqref{Majcondquant}, which gives the correct
form for the total energy~\eqref{totenquant} and the total
momentum~\eqref{totmomquant}.

Third, in Sec.~\ref{MATTEMF}, we have applied the elaborated
quantization method for the nondiagonal interaction with matter
and an electromagnetic field~\eqref{Hamint}. Within the forward
scattering approximation we have derived the quantized interaction
Hamiltonian~\eqref{Hamintquant}-\eqref{Fab}, which is valid for
arbitrary neutrino masses, an initial momentum, and the diagonal
neutrino interaction with matter. Then, using the density matrix
formalism, developed in Ref.~\cite{SigRaf93}, and in the
approximation of ultrarelativistic particles we have re-derived
the effective Hamiltonian~\eqref{Heffqm}, previously obtained in
frames of the standard quantum mechanical
approach~\cite{LimMar88}, for the description of spin-flavor
oscillations of Majorana neutrinos in matter and a magnetic field.

Finally, in Sec.~\ref{SI}, we have quantized the
self-interaction~\eqref{HSpsi} of Majorana neutrinos using the
developed formalism. Again in the forward scattering approximation
we have got the contribution to the effective
Hamiltonian~\eqref{HSquant} and~\eqref{MNKL} which is valid for
neutrinos with arbitrary masses and an initial momentum. Then, for
ultrarelativistic particles we have compared our results with the
previously obtained effective Hamiltonian~\cite{SigRaf93}, which
describes collective neutrino oscillations, and have found the
consistency.

Note that in all the previous works where collective effects in
neutrino flavor oscillations were studied the case of Dirac
neutrinos in a background matter was considered. Thus in the
present work for the first time we have discussed the situation of
Majorana neutrinos and generalized the consideration to include an
external magnetic field. It should be noted that for supernova
neutrinos both the interaction with a dense background matter, a
strong magnetic field, and the neutrino self-interaction can be of
equal importance. Therefore, the effective
Hamiltonians~\eqref{Heffqm}, \eqref{HSqm}, and~\eqref{HSqmcomp},
derived in our work, may be used for the treatment of collective
effects in spin-flavor oscillations of supernova neutrinos in
matter and a magnetic field.

The wave functions~\eqref{etaxisol} exactly take into account the
diagonal neutrino interaction with background matter, $g^0_{aa}$,
whereas in Sec.~\ref{MATTEMF} the external fields, $g^0_{ab}$, $a
\neq b$, and $F_{\mu\nu}$ are treated perturbatively, with only
linear term being kept, cf. Eqs.~\eqref{Mab} and~\eqref{Fab}.
However in general case the potentials $g^0_{aa}$ and $g^0_{ab}$
can be of the same order of magnitude. Nevertheless, besides
supernova neutrinos discussed here, we may apply the obtained
results to study nonperturbative effects in neutrino oscillations
if the nondiagonal potential of matter interaction $g^0_{ab}$ is
small or negligible. For instance, such a situation is implemented
when neutrinos propagate in the inner crust of a neutron star,
where $n_{e,p} \ll n_n$, with $n_{e,p,n}$ being the number
densities of electrons, protons, and neutrons respectively.
Another example, when the nondiagonal matter interaction is
unimportant, is $\nu_\mu \leftrightarrow \nu_\tau$ oscillations
channel.

\begin{acknowledgments}
I am very thankful to A.E.~Lobanov, G.G.~Raffelt, and V.B.~Semikoz
for helpful discussions as well as to FAPESP (Brazil) for a grant.
\end{acknowledgments}

\end{document}